\documentclass[12pt]{iopart}

\usepackage{graphicx}
\usepackage{color}

\def\spose#1{\hbox to 0pt{#1\hss}}
\def\simlt{\mathrel{\spose{\lower 3pt\hbox{$\mathchar"218$}}
     \raise 2.0pt\hbox{$\mathchar"13C$}}}
\def\simgt{\mathrel{\spose{\lower 3pt\hbox{$\mathchar"218$}}
     \raise 2.0pt\hbox{$\mathchar"13E$}}}
\begin{document}


\title[Dark matter: A phenomenological existence proof]{Dark matter: 
A phenomenological existence proof}

	\author{\textcolor{blue}{D  V  Ahluwalia-Khalilova}}

	\address{ ASGBG, Apartado Postal C-600, Department of Mathematics\\
   	University of Zacatecas (UAZ), Zacatecas 98060, Mexico}

	\ead{ahluwalia@heritage.reduaz.mx}
	     

\thispagestyle{empty}

\begin{abstract}
The non-Keplerian galactic rotational curves and the gravitational
lensing data~\cite{Persic:1995ru,Wittman:2000tc,Hoekstra:2003pn} 
strongly indicate a significant dark matter component
in the universe. Moreover, these data can be
combined to deduce  
the equation of state of dark matter \cite{Faber:2005xc}.
Yet, the existence of dark matter has been challenged
following the tradition of critical scientific spirit.
In the process, the theory of general relativity itself has been
questioned and various modified theories of gravitation have been
proposed~\cite{Milgrom:1983pn,Sanders:2002pf,Bekenstein:2004ne}. 
Within the framework of the 
Einsteinian general relativity,
here I make the observation that if the universe is
described by a
spatially flat Friedmann-Robertson-Walker (FRW)
cosmology with Einsteinian cosmological
constant
then the resulting cosmology predicts a 
significant dark matter component in the 
universe. The phenomenologically motivated existence 
proof refrains from invoking the
 data on galactic rotational curves and gravitational lensing,
but uses as input the age of the universe as deciphered from
studies on globular clusters. 
\end{abstract}

\textcolor{blue}{\hrule}


The abstracted result arises from the crucial observation
that in general, without invoking physics of structure formation (i.e.
primordial birth of galaxies and stars), it is impossible
to constrain the ratio of the fractional matter and dark
energy densities. However, an exception occurs, for the 
spatially flat FRW cosmology with Einsteinian cosmological
constant. In the `matter-dominated' epoch with
cosmological constant,
this exception completely determines the ratio
of the matter and dark energy densities as a function of the
age of the universe. Once the latter is brought in  as an input, the
said ratio is unambiguously fixed for the present epoch. It turns out that the 
resulting fractional density for matter $\Omega_\mathrm{m}$
is too large by a factor of 4 to 7 than the known 
$\Omega_\mathrm{sm} \approx 0.05$ associated with the
standard model of particle physics. This circumstance then unambiguously
predicts  a significant non-standard model component to $\Omega_\mathrm{m}$.
It is by rigorously establishing this outline in detail 
that the said claim is arrived at. 

No uniqueness
of the mentioned exception is implied; that is, there may exist
other cosmologies where a similar circumstance occurs.
I display the Newtonian constant $G$ explicitly but set the speed
of massless particles in vacuum $c$ to be unity. 

The setting of our argument is as follows.
I consider the matter-dominated epoch with Einsteinian cosmological
constant. By the latter one means  that  the dark energy equation
of state
$
      w^\Lambda = {p^\Lambda}/{\rho^\Lambda} 
$
is confined to  the choice
$w^\Lambda=-1$. Then,  a time independent  $\rho^\Lambda$ 
corresponds to the Einsteinian  cosmological constant.
Further, I consider the cosmology which is 
spatially flat. That is, spatial curvature constant $k=0$. 
In the matter dominated epoch, I set the associated pressure to
zero, $p=0$. This is quite realistic. Thus, the FRW cosmology
we consider is defined by the set:

\begin{equation}
\{k=0,\; w^\Lambda=-1,\; \rho=\rho_\mathrm{m},
\;p=p_\mathrm{m}=0,\;\;\rho^\Lambda={\mathrm{constant}}\}\label{eq:set}
\end{equation}

Once the  energy densities, $\rho_\mathrm{m}$ and $\rho^\Lambda$,
and the pressures, $p_\mathrm{m}$ and $p^\Lambda$,  associated with matter 
and cosmological constant $\Lambda$ are specified the 
scale factor $a(t)$, parameterizing the cosmological 
expansion, is obtained by solving
the Einstein field equations for
a spatially flat FRW cosmology:

\begin{eqnarray}
      \left(\frac{\dot{a}}{a}\right)^2 = \frac{8\pi G}{3}\rho_\mathrm{m} + 
      \frac{8\pi G}{3} \rho^\Lambda  
      \label{eq:1} \\[1ex]
       \frac{\ddot{a}}{a} = -\frac{4\pi G}{3} \rho_\mathrm{m}  - 
      \frac{4\pi G}{3}\left(\rho^\Lambda + 3 p^\Lambda\right) \,.
      \label{eq:2}
\end{eqnarray}
In writing the above equation
I set $p_\mathrm{m}=0$ for the matter dominated epoch,
and defined

\begin{equation}
\rho^\Lambda:= \frac{\Lambda}{8 \pi G}\,. \label{eq:rhoml}
\end{equation}
 Equivalently,
one can solve Eq.~(\ref{eq:1}), and the
equation for the conservation  of energy-momentum (which follows
from  Eqs.~(\ref{eq:1}) and (\ref{eq:2}))

\begin{equation}
      \dot{\rho}_\mathrm{m} + 3\left(\frac{\dot{a}}{a}\right)
      \rho_\mathrm{m} +
      3\left(\frac{\dot{a}}{a}\right)\left(\rho^\Lambda + p^\Lambda\right)
	 = 0\,.
      \label{eq:t3} 
\end{equation}
In obtaining the above equation 
$\dot{\rho}^\Lambda$ was set to zero as required by the setting,
and where in addition the last term on the left hand side now
vanishes due to the fact that $w^\Lambda= -1$ ($p^\Lambda=-\rho^\Lambda$), 
giving 

\begin{equation}
      \dot{\rho}_\mathrm{m} + 3\left(\frac{\dot{a}}{a}\right)
      \rho_\mathrm{m}  
	 = 0\,.
      \label{eq:4} 
\end{equation}

The system of Eqs.~(\ref{eq:1}) and  (\ref{eq:4}) 
now forms a closed system. It governs the time evolution
of $a(t)$ and $\rho(t)$.

For the spatially flat FRW cosmology defined by the set (\ref{eq:set}) 
the
division of Eq.~(\ref{eq:1}) by the square of the
Hubble parameter $H=\dot{a}/a$ yields

\begin{equation}
      1 = \Omega_{\mathrm m} + \Omega_{\Lambda}  \label{eq:one}
\end{equation}
where 

\begin{equation} 
      \Omega_{\mathrm m} :=\frac{8\pi G \rho_\mathrm{m}}{3 H^2} \,, \qquad
      \Omega_\Lambda := \frac{8 \pi G \rho^\Lambda}{3 H^2}
\stackrel{Eq.~(\ref{eq:rhoml})}{=}
\frac{\Lambda}{3 H^2} \,.
      \label{eq:t4}
\end{equation}
Our task now is to obtain an expression for the time
evolution of the ratio, $\Omega_\mathrm{m}:\Omega_\Lambda$.
Towards this end Eq.~(\ref{eq:4}) gives the temporal 
dependence of the matter density on the cosmic
scale factor

\begin{equation}
\rho_\mathrm{m}(t) =   \rho_\mathrm{m}(t_\star)
\left( \frac{a(t_\star)}{a(t)} \right)^3\, .\label{eq:rhom}
\end{equation}
Here, $t_{\star}$ marks the
beginning of the matter dominated epoch. It is to be given as
a physical input ($t_\star \ge 10^6~\mbox{years}$, temperature
$T \approx 2000~\mbox{K}$).

The time dependence of the scale factor $a(t)$
is now obtained by substituting for $\rho_\mathrm{m}(t)$ 
from Eq.~(\ref{eq:rhom}) into Eq.~(\ref{eq:1}),
and solving the resulting differential equation for $a(t)$.
This exercise yields the time evolution of the scale factor

\begin{equation}
      a(t) = a (t_\star) \left[ \left(\frac{\rho_m(t_\star)}{\rho^\Lambda}
\right)^{1/3} \sinh^{2/3} 
\left(\frac{\sqrt{3 \Lambda}}{2} t \right)\right] \,.
      \label{eq:a}
\end{equation}
This result agrees with that arrived at, e.g., by Frieman in 
Ref.~\cite[Eq.~3.6]{Frieman:1994gf}.\footnote{Padmanabhan 
\cite[Eq.~27]{Padmanabhan:2002ji} and Sahni \cite[Eq.~12]{Sahni:1999gb}
also note that scale factor $a(t)$ is proportional to $\sinh^{2/3} 
\left[(\sqrt{3 \Lambda}/2) t \right]$. However, as we next note, 
for the purposes of
establishing the proposed thesis it is precisely the 
`proportionality constant' that turns out to carry a pivotal
importance.}
 
From a physical perspective, note that the nonlinearity
of the gravitational field equations has the consequence
that the
``amplitude'' of the scale factor 
carries a 
unique dependence on the indicated
densities. Furthermore,  Eq.~(\ref{eq:a})  places
a severe constraint on the relative initial densities
by requiring the square bracket in Eq.~(\ref{eq:a}) to take  
the value unity  at $t=t_\star$:

\begin{figure}
\label{Fig:Zeta}
\noindent
\includegraphics{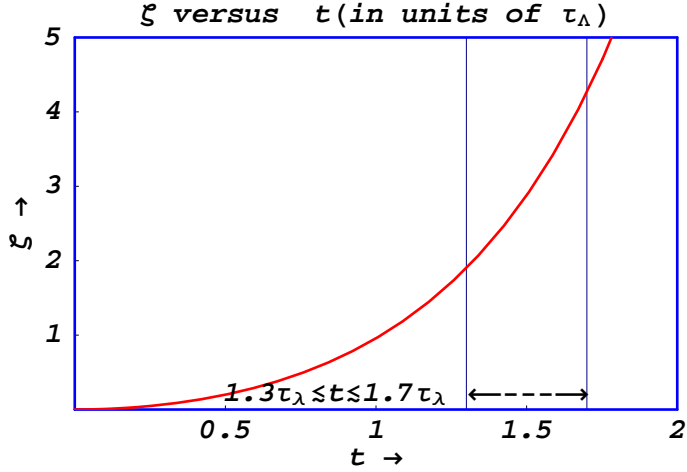}
\caption{The temporal 
evolution of $\zeta:=\Omega_\Lambda/\Omega_\mathrm{m}$.
The label $\leftrightarrow$ marks the present epoch
with $t_0 \approx (13.5 \pm 1.5)\times
10^{9}$ years in units of $\tau_\Lambda$ ($=9 \times 10^{9}$ years,
see text).}
\end{figure}

\begin{equation}
\frac{\rho_\mathrm{m}(t_\star)}{\rho^\Lambda} = 
\left[\sinh
\left(\frac{\sqrt{3}}{2} \frac{t_\star}{\tau_\Lambda}\right) 
\right]^{-2}\label{eq:constraint} 
\end{equation}
where we introduced $\tau_\Lambda:=\sqrt{1/\Lambda}$.
Substituting back $a(t)$ from Eq.~(\ref{eq:a}) 
in Eq.~(\ref{eq:rhom})
provides the general time evolution
of the relevant energy densities for $t\ge t_\star$

\begin{equation}
\frac{\rho_\mathrm{m}(t)}{\rho^\Lambda} = 
\left[\sinh
\left(\frac{\sqrt{3}}{2} \frac{t}{\tau_\Lambda}\right) \right]^{-2}\,.\label{eq:12}
\end{equation}
The result (\ref{eq:constraint}) also follows from the above 
equation for $t=t_\star$. It reflects the internal consistency
of the calculations.
On dividing each of the densities in Eq.~(\ref{eq:12}) by the critical density,
$\rho_c:=3 H^2/(8 \pi G)$,
the  time evolution
of the ratio $\Omega_\mathrm{m}:\Omega_\Lambda$ follows

\begin{equation}
\Omega_\mathrm{m}(t) : \Omega_\Lambda(t) = 1:\zeta(t)\label{eq:zeta}
\end{equation}
where
$
\zeta(t):= \sinh^2
\left(\sqrt{3 }\, t/ (2 \tau_\Lambda) \right).
$
Once the age of the universe is specified by some 
independent observations,  
the considered cosmology uniquely determines
the ratio $\Omega_\mathrm{m} : \Omega_\Lambda$ and 
it in addition predicts the fractional matter density to be

\begin{equation}
\Omega_\mathrm{m}(t)= (1+\zeta(t))^{-1}\,.
\end{equation}
This is the combined result of Eqs.~(\ref{eq:one}) and (\ref{eq:zeta}).

Since 1998, there exists observational evidence for
a positive cosmological constant  
\cite{Perlmutter:1997zf,Riess:1998cb,Perlmutter:1998np,Tonry:2003zg,Riess:2004nr}.
The latest data and analysis of the luminosity 
data on Supernovae 1a, in addition, obtains \cite{Astier:2005qq}
(for a $k=0$ cosmology)  

\begin{figure}
\label{Fig:Omega}
\noindent
\includegraphics{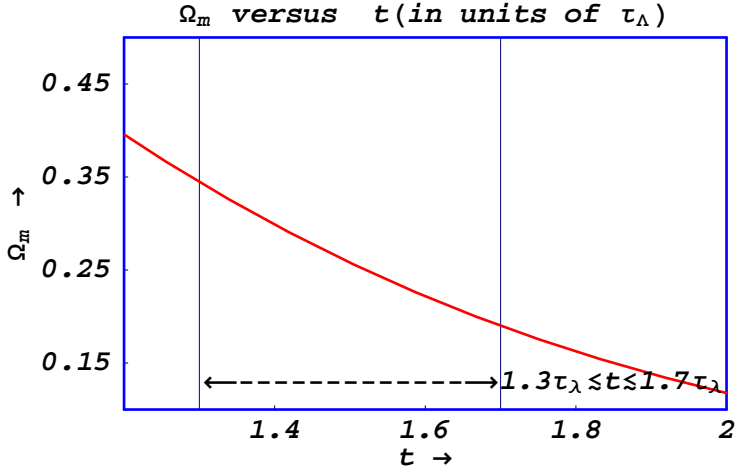}
\caption{The temporal 
evolution of $\Omega_\mathrm{m}$.
The label $\leftrightarrow$ marks the present epoch
with $t_0 \approx (13.5 \pm 1.5)\times
10^{9}$ years in units of $\tau_\Lambda$ ($=9 \times 10^{9}$ years,
see text).
}
\end{figure}

\begin{equation}
w^\Lambda= -1.023 \pm 0.090 \;(stat)\pm 0.054\;(syst)\,.\label{eq:wlambda}
\end{equation}
These observations give us confidence that a spatially flat
FRW cosmology with Einsteinian cosmological constant may indeed
correspond to the physical reality. 
Taking $\Lambda/(8\pi G)
\approx 4 \times 10^{-47}~\mbox{GeV}^4$ yields $\tau_\Lambda \approx 
9 \times 10^{9}$ years. It sets the time scale for us.\footnote{Indirect
evidence for dark energy also comes from the data on the
cosmic microwave background
and the large scale structure observations \cite{Spergel:2003cb,Tegmark:2003ud}.}

However, for the moment, one need not take
any input from these data, except for working
in units of $\tau_\Lambda$ (for convenience). Instead, I use 
the present age of the universe $t_0 \approx (13.5 \pm 1.5)\times
10^{9}$ years \cite{Chaboyer84:2001} as determined by the studies on 
globular clusters as the input. 
Thus the present epoch corresponds roughly to the range:
$1.3 \tau_\Lambda \le t \le 1.7 \tau_\Lambda$. 
In this range, Figures 1 and 2 
show  that 

\begin{equation}
1.9 \le \zeta \le 4.3\,,\qquad
0.19 \le \Omega_\mathrm{m} \le 0.35 \,.\label{eq:prediction}
\end{equation}
The latter may be compared
with $\Omega_\mathrm{m} = 0.271 \pm 0.021 (stat) \pm 0.007 (syst)$
inferred by Astier~et~al.~\cite{Astier:2005qq}.
However, $\Omega_\mathrm{sm} \approx 0.05$
for the present epoch
can only account for a small
fraction, i.e., one fourth to one seventh  
of the predicted $ \Omega_\mathrm{m}$ as given in
 Eq.~(\ref{eq:prediction}). 
This state of affairs
leaves the fractional density $\Omega_\mathrm{m}-\Omega_\mathrm{sm}$
to point towards the existence of
some form of non-standard model matter in the
non-relativistic form. This, by definition, is the astrophysical/cosmic 
dark matter,  with
fractional density given by

\begin{figure}
\label{Fig:THubble}
\noindent
\includegraphics{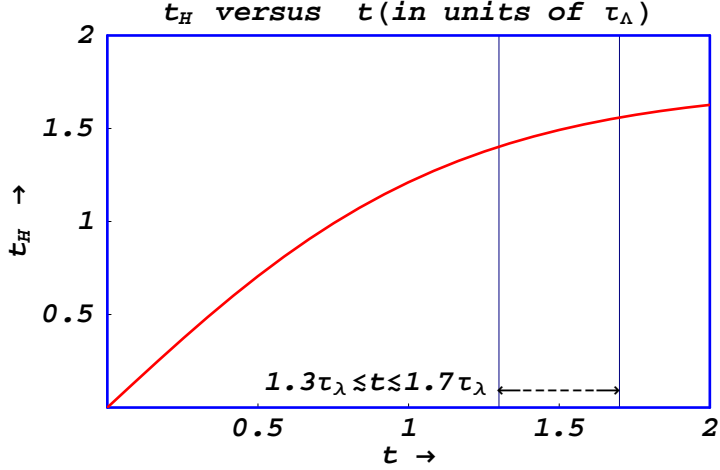}
\caption{
The range $1.3 \tau_\Lambda \le t \le 1.7 \tau_\Lambda$ corresponds
to   $1.4 \tau_\Lambda \le t_H \le 1.6 \tau_\Lambda$. The $t_H$ is
defined as inverse of the Hubble parameter for spatially
flat FRW cosmology with Einsteinian cosmological constant.}
\end{figure}

\begin{equation}
\Omega_\mathrm{dm}(t) = \Omega_\mathrm{m}-\Omega_\mathrm{sm} =
(1+\zeta(t))^{-1}  - \Omega_\mathrm{sm}(t)\,.
\end{equation}
For the present epoch, it yields
roughly: $0.14 \le \Omega_\mathrm{dm} \le 0.30$.

In the defined setting,
the $\Omega_\mathrm{dm}$ as well as the  $\Omega_\mathrm{sm}$
carry the same temporal evolution. It is immediately read off
from  Eqs.~(\ref{eq:rhom}) and (\ref{eq:a}). In particular, the
ratio  $\Omega_\mathrm{dm}: \Omega_\mathrm{sm}$ is frozen 
in time.

Before concluding it may be noted that the Hubble
parameter as implied by the obtained scale factor 
given in Eq.~(\ref{eq:a}) may also be evaluated. The result is,

\begin{equation}
H(t):\frac{\dot{a}}{a} = \sqrt{\frac{\Lambda}{3}}
\coth\left( \frac{\sqrt{3}}{2}\frac{t}{\tau_\Lambda}\right)\,.
\end{equation}
It defines a characteristic time scale which is
closely related to the age of the universe

\begin{equation}
t_H := \frac{1}{H(t)}=\sqrt{3}  
\tanh\left( \frac{\sqrt{3}}{2}\frac{t}{\tau_\Lambda}\right)
\tau_\Lambda \,.
\end{equation}
The range $1.3 \tau_\Lambda \le t \le 1.7 \tau_\Lambda$ corresponds
to   $1.4 \tau_\Lambda \le t_H \le 1.6 \tau_\Lambda $ (see
Fig. 3). A more rigorous analysis along the lines of Ref. 
\cite[Sec. 3.2]{KolbTurner:1990}
yields a very similar result.

Despite the on going debate on whether or not the general 
theory of relativity needs a modification while confronting
the astrophysical and cosmological scene, the phenomenological
existence
proof given in the preceding discussion makes a strong case
that such modifications are not required by the data 
on the galactic rotational curves or gravitational lensing.
The dark matter required by these latter observations is
the same dark matter that is here predicted within the general
relativistic framework for the
observationally favored spatially flat
FRW cosmology with Einsteinian cosmological constant.
The tantalizing possibility exists that 
the Pioneer anomaly may require a modification of the 
general theory of relativity in a 
subtle way~\cite{Turyshev:2005ej,Grumiller:2006privatecommunication}.
But that call comes not from the dark matter.
The focus on new physics possibilities that dark matter 
offers perhaps ought to shift dramatically. One such direction
is provided by the  new and entirely unexpected
theoretical discovery of a mass dimension one fermionic field
of spin one half \cite{Ahluwalia-Khalilova:2004sz,Ahluwalia-Khalilova:2004ab}.
That fermionic field, by virtue of its mass dimension, explains the darkness
of dark matter as a natural consequence of the spacetime symmetries
underlying the freely falling frames of general relativity.

\section*{Acknowledgments}

I wish to thank Daniel Grumiller for his reading of the final
draft, and for his input. I also thank Daniel Sudarsky for a remark
on a related work which led to the considerations presented here.

\textcolor{red}{\hrule}

\end{document}